# Specification and Verification of Uplink Framework for Application of Software Engineering using RM-ODP

Krit Salahddine[1], Laassiri Jalal[2] and El Hajji Said[3]

[1] Polydisciplinary Faculty of Ouarzazate, University Ibn Zohr, BP/638, Morocco

[2] Department of informatics, Faculty of Sciences, University Ibn Tofail, BP 33, Morocco

[3] Department of Mathematic Informatics, University Mohamed V-Agdal, BP 1040, Morocco

**Abstract**

This paper present a survey and discussion of the Reference Model for Open Distributed Processing (RM-ODP) viewpoints; oriented approaches to requirements engineering viewpoint and a presentation of new work in the application wireless mobile phone, this area which has been designed with practical application using the Unified Modelling Language (UML)/VHDL_AMS (VHSIC Hardware Description Language Analog and Mixed-Signal). We mainly focus on rising and fulling time, action, uplink behaviour constraints (sequentiality, non determinism and concurrency constraints).We discuss the practical problems of introducing viewpoint; oriented requirements engineering into industrial software engineering practice and why these have prevented the widespread use of existing approaches.

The goal of this article is to check the uplink path using the MIC (Microphone amplifier) with all analog inputs, and check the amplifier gain.

This paper provides an example of using the Uplink Framework to build a comprehensive, good solution for Application Wireless Mobile Phone.

Finally, we discuss how well this approach addresses some outstanding problems in requirements engineering (RE) and the practical industrial problems of introducing new requirements engineering methods.

## 1. Introduction

The rapid growth of distributed processing has led to a need for coordinating framework for the standardization of Open Distributed Processing (ODP). The Reference Model for Open Distributed Processing (RM-ODP) [1]-[4] provides a framework within which support of distribution, networking and portability can be integrated. The foundations part [2] contains the definition of the concepts and analytical framework for normalized description of (arbitrary) distributed processing systems. These concepts are grouped in several categories. The architecture part [3] contains the specifications of the required characteristics that qualify distributed processing to be open. It defines a framework comprising five viewpoints, viewpoint language, ODP functions and ODP transparencies. The five viewpoints, called enterprise, information, computational, engineering and technology provide a basis for the specification of ODP systems.

Each viewpoint language defines concepts and rules for specifying ODP systems from the corresponding viewpoint. The ODP functions are required to support ODP systems.

In this context, VHDL_AMS is used to specify the properties to be tested. The UML meta-models provide a precise core of any ODP tester. We use in this paper ModelSim under Cadence to verify process behavior based on interaction and the binding object in the ODP systems.

VHDL_AMS is an industry standard modeling language for mixed signal circuits. It provides both continuous-time and event-driven modeling semantics, and so is suitable for analog, digital, and mixed analog/digital circuits. It is particularly well suited for verification of very complex analog, mixed-signal and radio frequency integrated circuits.

This capability is used to highlight some benefits of the Architectural realized (uplink path): raising the level of abstraction at which development occurs; which, in turn, will deliver greater productivity, better quality, and insulation from underlying changes in technology.

We treated the need of formal notation for Uplink behavioral concepts in the Computational language [8]. Indeed, the viewpoint languages are abstract in the sense that they define what concepts should be supported, not how these concepts should be represented. It is important to note that, RM-ODP uses the term language in its broadest sense: "a set of terms and rules for the construction of statements from the terms". It does not propose any notation to support the viewpoint languages. Using the Unified Modeling Language (UML)/VHDL_AMS [9], [10] we defined a formal semantic for a fragment of ODP uplink behavior concepts defined in the RM-ODP foundations part and in the engineering language [11]. These concepts (time, action, uplink behavior constraints) are suitable for describing and constraining the uplink behavior of ODP engineering viewpoint specifications.

## 2. Meta-MODELING Time and Behavioral Constraints

Behavioral constraints may include sequentiality, non-determinism, concurrency, real time" (RM-ODP, part 2, clause 8.6). In this work we consider constraints of sequentiality, non-determinism and concurrency. The concept of constraints of sequentiality is related with the concept of time.

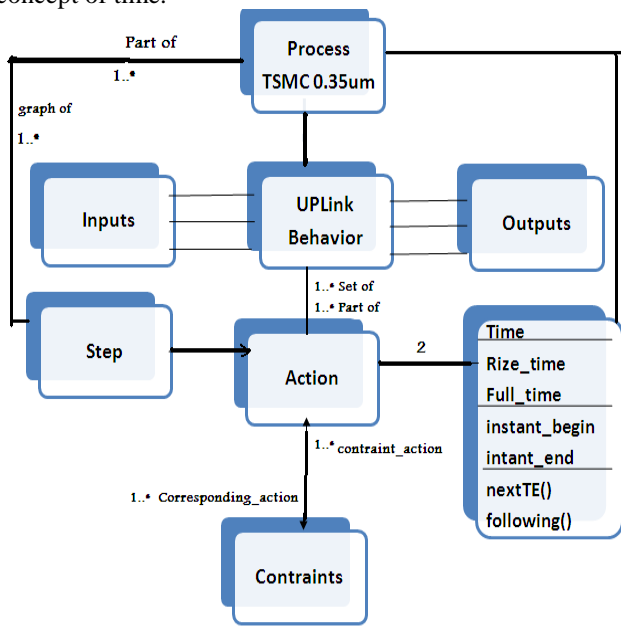

Fig.1 Meta-Modelling core Behavior concept in RM-ODP foundations part

### 2.1 Time

Time has two following important roles in system design [26]:

•It serves for the purpose of synchronization of actions inside and between processes, the synchronization of a system with system users, the synchronization of user requirements with an actual performance of a system.
•It defines sequences of events (action sequences)

To fulfil the first goal, we have to be able to measure time intervals. However, a precise clock that can be used for time measurement does not exist in practice but only in theory [27]. So the measurement of the time is always approximate. In this case we should not choose the most precise clocks, but ones that explain the investigated phenomena in the best way. Simultaneity of two events or their sequentiality, equality of two durations should be defined in the way that the formulation of the physical laws is the easiest" [27]. For example, for the actions synchronization, internal computer clocks can be used and, for the synchronization of user requirements, common clocks can be used that measure time in seconds, minutes and hours.

We consider the second role of time. According to [27] we can build some special kind of clock that can be used for specifying sequences of actions. RM-ODP confirms this idea by saying that "a location in space or time is defined relative to some suitable coordinate system" (RM_ODP, part 2, clause 8.10). The time coordinate system defines a clock used for system modeling. We define a time coordinate system as a set of time events. Each event can be used to specify the beginning or end of an action. A time coordinate system must have the following fundamental properties [26]:
  •Time is always increasing. This means that time cannot have cycles.
  •Time is always relative. Any time moment is defined in relation to other time moments (next, previous or not related). This corresponds to the partial order defined for the set of time events.

We use the UML (fig1) and OCL to define time: Time is defined as a set of time events.
**nextTE**: defines the closest following time events for any time events [26].

We use the followingTE relation to define the set of the following time events or transitive closure for the time event t over the nextTE relation:
**followingTE**: defines all possible following time events Using followingTE we can define the following invariant that defines the transitive closure and guarantees that time event sequences do not have loops :
**Context** t : time  **inv** :
Time->forAll(t:Time | (t.nextTE->isempty  implies t.follwingTE->isempty)
and  (t.nextTE->notempty  and  t.follwingTE->isempty  implies t.follwingTE =t.nextTE)    and  (t.nextTE->notempty  and t.follwingTE->notempty         implies          t.follwingTE->

includes(t.nextTE.follwingTE->union(t.nextTE)) and t.follwingTE->exludes(t)).

This definition of time is used in the next section to define sequential constraints.

### 2.2 Behavioral constraints

We define the behavior like a finite state automaton (FSA). For example, figure 2 shows a specification that has constraints of sequentiality and non determinism. The system is specified using constraints of non-determinism since state S1 has a non-deterministic choice between two actions a and b.

Based on RM-ODP, the definition of behavior must link a set of actions with the corresponding constraints. In the following we give definition of constraints of sequentiality, of concurrency and of non-determinism.

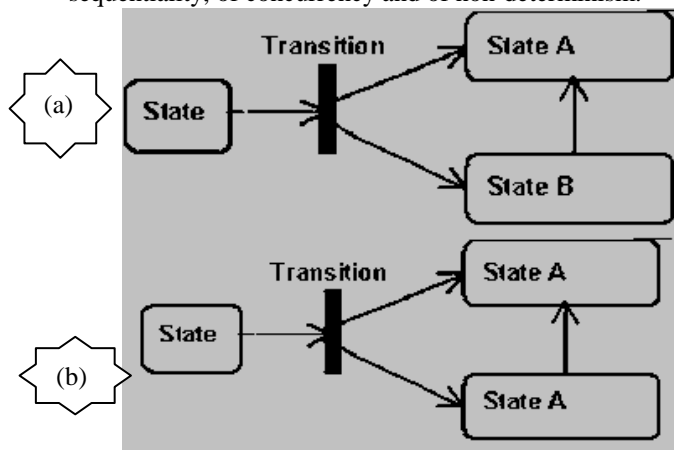

Fig. 2. a - Sequential deterministic constraints;

b - Sequential non deterministic constraints.

### B.1 Constraints of sequentiality

Each constraint of sequentiality should have the following properties [26]:
• It is defined between two or more actions.

• Sequentiality has to guarantee that one action is finished before the next one starts. Since RM-ODP uses the notion of time intervals it means that we have to guarantee that one time interval follows the other one:

**Context** sc : constraintseq **inv** :
Behavior.actions-> forAll(a1,a2 | a1<> a2 and a1.constraints->includes(sc)
and a2.constraints->includes(sc) and
((a1.instant_end.followingTE->includes(a2.instant_begin)
or(a2.instant_end.followingTE->includes(a1.instant_begin) )

For all SeqConstraints sc, there are two different actions a1, a2, sc is defined between a1 and a2 and a1 is before a2 or a2 is before a1.

### B.2 Constraints of concurrency

Figure 3 shows a system specification that has constraints of concurrency since state a1 has a simultaneous choice of two actions a2 and a3.

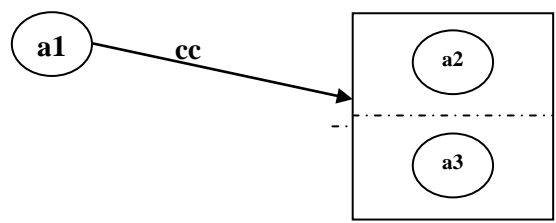

Fig. 3. RM-ODP diagram: Example constraints of concurrency

For all concuConstraints cc there is a action a1, there are two different internal actions a2, a3, cc is defined between a1 and a2 and a3, a1 is before a2 and a1 is before a3
**Context** cc: constraintconc inv:
Behavior.actions-> forAll(a1 :Action ,a2 ,a3 : internalaction | (a1 <> a2) and
(a2 <> a3) and (a3 <> a1) and a1.constraints->includes(cc) and a2.constraints->includes(cc) and a3.constraints->includes(cc) and
a1.instant_end.followingTE-> includes(a2.instant_begin) and
a1.instant_end.followingTE-> includes(a3.instant_begin))

### B.3 Constraints of non-determinism

In order to define constraints of non-determinism we consider the following definition given in [24]: "A system is called non-deterministic if it is likely to have shown number of different behavior, where the choice of the behavior cannot be influenced by its environment". This means that constraints of non-determinism should be defined between a minimum of three actions. The first action should precede the two following actions and these actions should be internal (see figure 4).

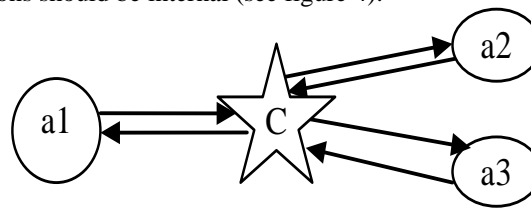

Fig. 4. Example Constraints example of non-determinism

We define this constraint as follows:
**Context** ndc: NonDetermConstraints **inv**:
Behavior.actions-> forAll(a1 :Action ,a2 ,a3 : internalaction | (a1 <> a2) and
(a2 <> a3) and (a3 <> a1) and a1.constraints->includes(ndc) and
a2.constraints->includes(ndc) and
a3.constraints->includes(ndc) and
a1.instant_end.followingTE-> includes( a2.instant_begin) or
a1.instant_end.followingTE-> includes(a3.instant_begin)) .

We note that, since the choice of the behavior should not be influenced by environment, actions a2 and a3 have to be internal actions (not interactions). Otherwise the choice between actions would be the choice of environment [26].

## 3. Simulation results and discussion

Simulations are carried on using VHDL-AMS after verification all of the connectivity between the different actions.

The goal of action 1 is to check the uplink path using the MIC amplifier with all analog inputs, and check the amplifier gain.

Sine waves (at different frequencies) are sent from all analog inputs to the VSIF through the uplink path (MIC amplifier, ADC).

First MICN/P differential input is selected Fig.5.

Then FML mono input is selected, and then HSMIC mono input is selected fig.6.

The MIC amplifier gain is set from 3dB to 33dB.

Checks are done on the MIC amp outputs and ADC outputs: signal, gain fig.7

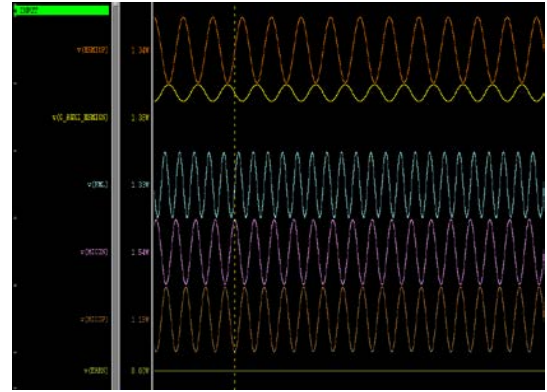

Fig.6 Input analog signals

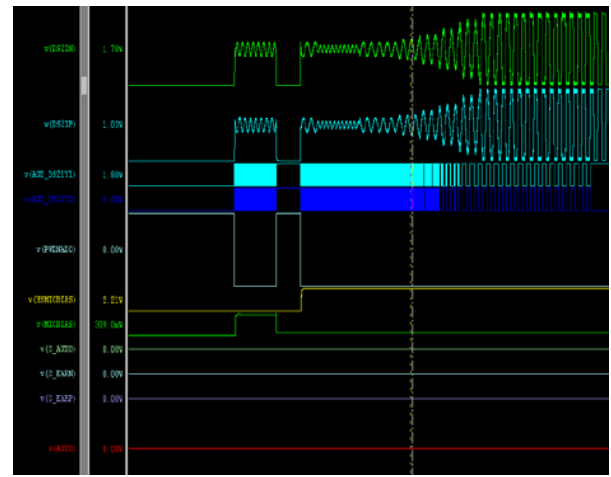

Fig.7 microphone amplifier signals

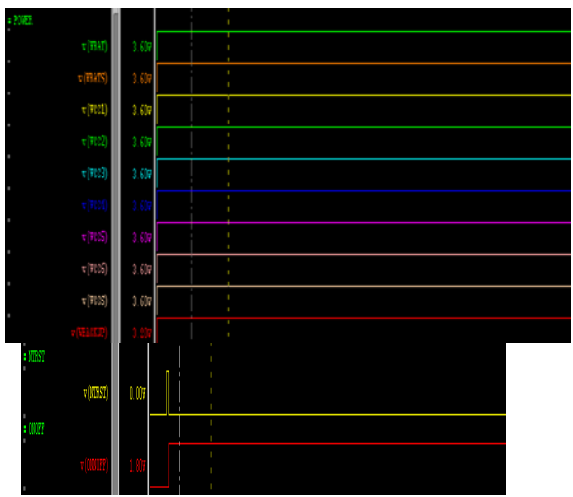

Fig.5 NPUT signals

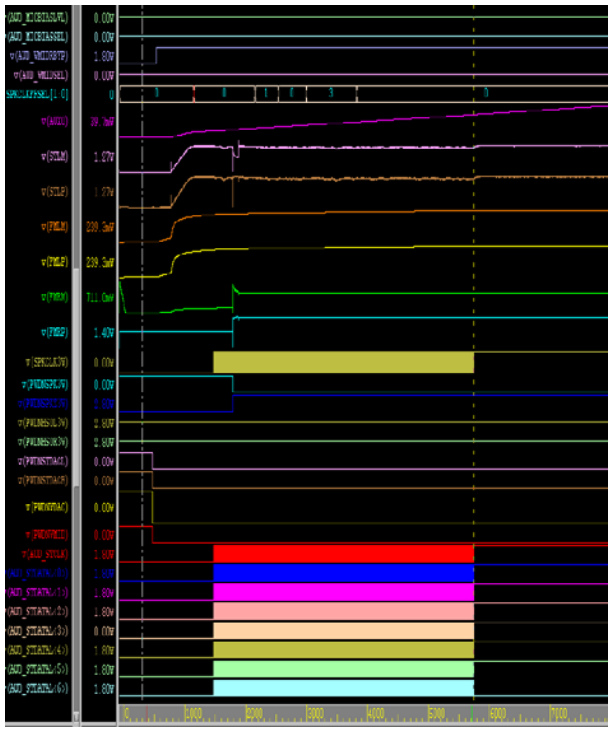

Fig.8 Speaker output

## 4. Conclusions

An Architectural Framework for Wireless Mobile has been proposed. This Architectural Framework will allow Uplink to be used in any wireless environment, as required, to provide any type of services demanded by the user regardless whether the uplink is a stand alone, a pure Wireless sensor network or integrated with other networks. In this paper we have presented our contribution to the RMODP standard-related research. This contribution resolves an important problem of the RM-ODP standard: the absence of a single consistent formalization of the RM-ODP conceptual framework. A realization of such formalization was officially verified.

The goal of our work is to help promote the practical applications of RM-ODP. The formal model of RM-ODP Part 2 that we presented in this paper can indeed serve for the promotion of RM-ODP towards a wider use in the modern modeling practices. Some of the applications of our results have already justified this claim.

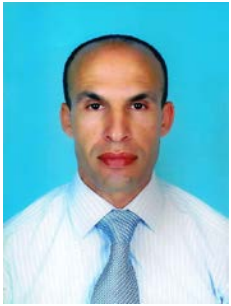

**Salah-ddine Krit** received the B.S. and Ph.D degrees in Microectronics Engineering from Sidi Mohammed Ben Abdellah university, Fez, Morroco. Institute in 2004 and 2009, respectively. During 2002-2008, he is also an engineer Team leader in audio and power management Integrated Circuits (ICs) Research.

 Design, simulation and layout of analog and digital blocks dedicated for mobile phone and satellite communication systems using CMOS technology. He is currently a professor of informatics with Polydisciplinary Faculty of Ouarzazate, Ibn Zohr university, Agadir, Morroco. His research interests include wireless sensor Networks (Software and Hardware), computer engineering and wireless communications.

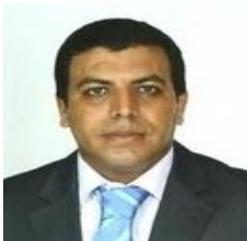

**Jalal Laassiri** received his Bachelor's degree (License es Sciences) in Mathematics and Informatics in 2001 and his Master's degree (DESA) in computer sciences and engineering from the faculty of sciences, university Mohammed V, Rabat, Morocco, in 2005, and he developed He received his Ph.D. degree in computer sciences and engineering from University of Mohammed V, Rabat, Morocco, in Juin, 2010. He was a visiting scientific with the Imperial College London, in London, U.K. He is Member of the International Association of Engineers (IAENG), He joined the Faculty of Sciences of Kénitra, Department of Computer Science , Ibn Tofail University, Morocco, as an Professor in October 2010, His current research interests include Software and Systems Engineering, UML-OCL, B-Method, ..